\documentclass[aps,prl,preprint,showpacs,superscriptaddress]{revtex4}
\usepackage{graphicx}
\usepackage{calc}
\usepackage{bm}

\begin{document}

\title{Electrical read-out of the local nuclear polarization \\
in the quantum Hall effect\\
-- A hyperfine battery --}

\author{A.~W\"urtz}
\altaffiliation[Present address: ]{ATMEL Germany GmbH, Theresienstr. 2,
D-74072 Heilbronn, Germany}
\affiliation{Laboratorium f\"ur Festk\"orperphysik, Universit\"at
Duisburg-Essen, Lotharstr. 1, D-47048 Duisburg, Germany}
\author{T.~M\"uller}
\affiliation{Laboratorium f\"ur Festk\"orperphysik, Universit\"at
Duisburg-Essen, Lotharstr. 1, D-47048 Duisburg, Germany}
\author{A.~Lorke}
\email[Corresponding~author, e-mail: ]{lorke@uni-duisburg.de}
\affiliation{Laboratorium f\"ur Festk\"orperphysik, Universit\"at
Duisburg-Essen, Lotharstr. 1, D-47048 Duisburg, Germany}
\author{D.~Reuter}
\affiliation{Lehrstuhl f\"ur Angewandte Festk\"orperphysik,
Ruhr-Universit\"at Bochum, Universit\"atsstr. 150, D-44780 Bochum,
Germany}
\author{A.~D.~Wieck}
\affiliation{Lehrstuhl f\"ur Angewandte Festk\"orperphysik,
Ruhr-Universit\"at Bochum, Universit\"atsstr. 150, D-44780 Bochum,
Germany}

\date{\today}

\begin{abstract}
It is demonstrated that the now well-established `flip-flop'
mechanism of spin exchange between electrons and nuclei in the
quantum Hall effect can be reversed. We use a sample geometry
which utilizes separately contacted edge states to establish a
local nuclear spin polarization --close to the maximum value
achievable-- by driving a current between electron states of different
spin orientation. When the externally applied current is switched
off, the sample exhibits an output voltage of up to a few tenths
of a meV, which decays with a time constant typical for the
nuclear spin relaxation. The surprizing
fact that a sample with a local nuclear spin polarization can act as a
source of energy and that this energy is well above the nuclear Zeeman
splitting is explained by a simple model which takes into account the
effect of a local Overhauser shift on the edge state reconstruction.

\end{abstract}


\maketitle

The recent interest in nuclear spin polarization in
semiconductor-based electronic structures arises mainly from their
possible application for quantum computing
\cite{q-comp}. As a consequence, experimental and
theoretical emphasis has been put on studies of nuclear spin
relaxation mechanisms and times \cite{berg,mozyrsky,bychkov} as
well as on the electrical controllability of the nuclear spin
polarization \cite{hashimoto,johnson}. The interaction between the
spins of electrons and nuclei have been proposed as a
mechanism for the write-in process in quantum information storage
and processing. On the other hand, the influence of the nuclear
spin polarization on electronic transport properties via the
hyperfine interaction may provide a possible read-out process
\cite{dobers,desrat}. The device presented in the following
combines both, write-in and read-out of the spin state stored in
the nuclear system using transport between spin-polarized quantum
Hall edge states.

One way to locally manipulate nuclear spins is to make use of the
spin-polarized channels formed at the edges of a two-dimensional
electron system (2DES) in the quantum Hall regime. The spin-flip
associated with transport between two edge states of different
spin orientation can be mediated by a reverse `spin-flop' in
the nuclear spin system, which over time leads to a built-up of a
nuclear spin polarization \cite{berg,dobers}.
A number of experiments have demonstrated such electrically pumped
nuclear spin polarization and how it affects electronic transport
in high magnetic fields \cite{dixon,wald,machida,edik,smet}.

Here we introduce an experimental approach which makes use of
separately contacted, spin-polarized edge channels and allows us not only to
induce a local nuclear spin polarization at a predefined site
\cite{edik}. It also provides for a simple and accurate means to
determine the local hyperfine field. We find that the non-equilibrium
state associated with the spatial variation of the nuclear
polarization can serve as an internal source of energy. So the
nuclear system in the sample, after being energized by the
flip-flop mechanism, can be viewed as a `hyperfine battery'. By
measuring the voltage between the edge channels, we can directly determine
the Overhauser shift as a function of time. We find clear evidence that
the flip-flop
mechanism is most effective for establishing a nuclear spin polarization
when the electrons are transferred resonantly, i.e., when the applied
bias corresponds to the electronic Zeeman gap.

\begin{figure}
\includegraphics*[width=\columnwidth-4cm]{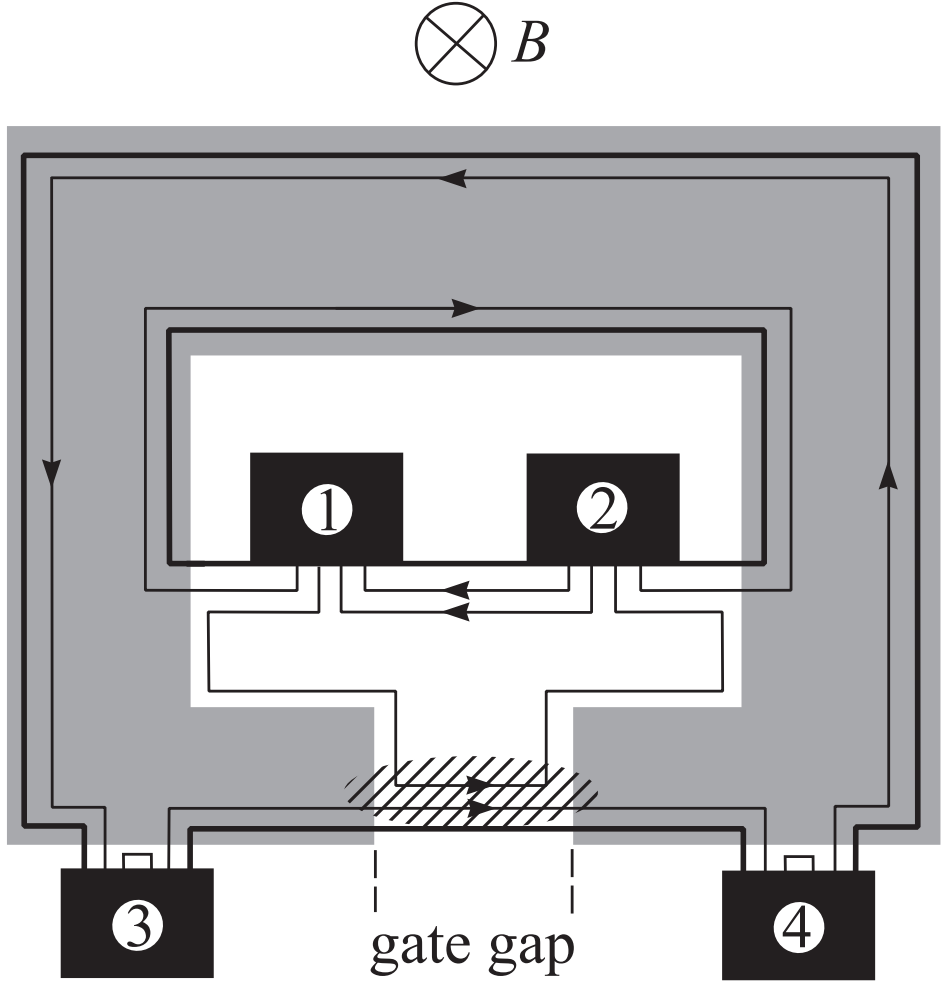}
\caption{Schematic of the sample geometry. Contacts (1--4) are
positioned along the etched edges (thick lines) of the ring-shaped mesa. The
grey area represents the gate electrode, the hatched area indicates the 
interaction region of the edge states, where the nuclear spin polarization is induced. 
Arrows indicate the
direction of electron drift in the ESs for the chosen magnetic
field direction. The filling factors are $\nu=2$ in the ungated
regions and $g=1$ under the gate. The gate gap width is
$2\:\mathrm{\mu m}$. \label{fig1}}
\end{figure}

The samples are fabricated from two different
$\mathrm{GaAs/}\mathrm{Al}_{x}\mathrm{Ga}_{1-x}\mathrm{As}$
heterostructures (A and B), grown by molecular-beam epitaxy. They contain a
2DES, located 70 nm and 110 nm below the surface, respectively. The
mobilities and carrier densities at $4.2\:\mathrm{K}$ are
$8\times 10^{5}\:\mathrm{cm}^{2}/\mathrm{Vs}$ and $3.7\times
10^{11}\:\mathrm{cm}^{-2}$ for sample A,
$2.2\times 10^{6}\:\mathrm{cm}^{2}/\mathrm{Vs}$ and
$1.35\times 10^{11}\:\mathrm{cm}^{-2}$ for sample B. Different samples prepared from
both heterostructures exhibit the same behavior. For consistency, only
results from sample A are presented in the following.

To separately contact single edge channels of different spin
orientation, we use an etched, quasi-ring-shaped mesa geometry
\cite{alida,muller}, as shown in Fig.
\ref{fig1} (thick lines). Two Ohmic contacts (1 and 2) are located in the
interior, two others (3 and 4) on the outside of the mesa. A
horseshoe-shaped gate electrode (shaded grey) covers most of the
sample, leaving uncovered only the narrow gate gap region on the
outer sample edge, which determines the location where the nuclear
spin polarization will be induced and detected. An external magnetic
field is applied so that the filling factor
$\nu$ in the ungated regions of the sample is $\nu=2$
($\nu=n/n_{B}$, with $n$ being
the electron density and $n_{B}=eB/h$ the degeneracy of each
Landau level). A negative voltage is applied to the gate
electrode to adjust the local filling factor under the gate
to $g=1$.

Electronic transport between the inner (1, 2) and
outer contacts (3, 4) is only possible in the gate gap
region, where the edge channels are running parallel in close proximity. For
the chosen filling factor combination, the edge channel connected to contacts
1 and 2 is spin-down (antiparallel to the external magnetic
field), whereas the one connected to contacts 3 and 4 is spin-up
(parallel to the external magnetic field). This way,
four-probe transport measurements are possible between
spin-polarized electron states. For a more detailed
description of the sample topology and the transport spectroscopy of
separately contacted edge channels, see \cite{alida,edik}. It should be
noted that even though a number of other sample layouts are also
suitable for studying hyperfine interactions in the quantum Hall regime
\cite{dixon,wald,machida},
the present topology gives a very high impedance \cite{alida,muller}, which enables the
measurements below.

The experiments shown here
are performed in a $^{3}\mathrm{He}$-cryostat at a temperature of
$T=240\:\mathrm{mK}$ and a magnetic field of $B =
9.1\:\mathrm{T}$. First, a stable nuclear polarization is
established in the gate gap region by driving a constant current
$|I_{pump}| \leq 500\:\mathrm{nA}$ between the edge states in the
gate gap. It is well established \cite{dobers,berg,dixon,wald} that charge
transfer between electron reservoirs of opposite spin requires
an electronic `spin-flip' which can be mediated by a
nuclear `spin-flop' in the opposite direction. Because nuclear spin relaxation
times are typically several tens to thousands of
seconds, a
net nuclear spin polarization builds up at the location where the
electron spin-flip is induced. This process of `dynamic nuclear
polarization' (DNP) will in turn affect the
energetic structure of the electron system through the hyperfine
interaction. The Hamiltonian of the hyperfine interaction is given
by
\begin{equation}
A\bm{I \cdot S}=\frac{1}{2}A(I^+S^- + I^-S^+)+ A I_{z} S_{z}
\nonumber\quad ,
\end{equation}
where $A>0$ is the hyperfine constant, $\bm{I}$ is the nuclear and
$\bm{S}$ is the electron spin. The final term accounts for the
contact-hyperfine interaction between the z-components of the nuclear
and the electron spin ($I_{z}$ and $S_{z}=\pm 1/2$), respectively.

In general, a stable nuclear polarization is established within a few
minutes \cite{dixon,wald}. In the present setup, this can be confirmed
\textit{in-situ} by 4-probe measurements of
the resistance between inner and outer contacts \cite{edik}. To
polarize the nuclear spin system, we apply $I_{pump}$ for 10 minutes,
until a constant voltage drop has established between inner and outer
contacts. This procedure provides for a
reproducible initial state of the system with a nuclear
polarization, depending on the sign and magnitude of $I_{pump}$
\cite{edik}.

\begin{figure}
\includegraphics[width=\columnwidth]{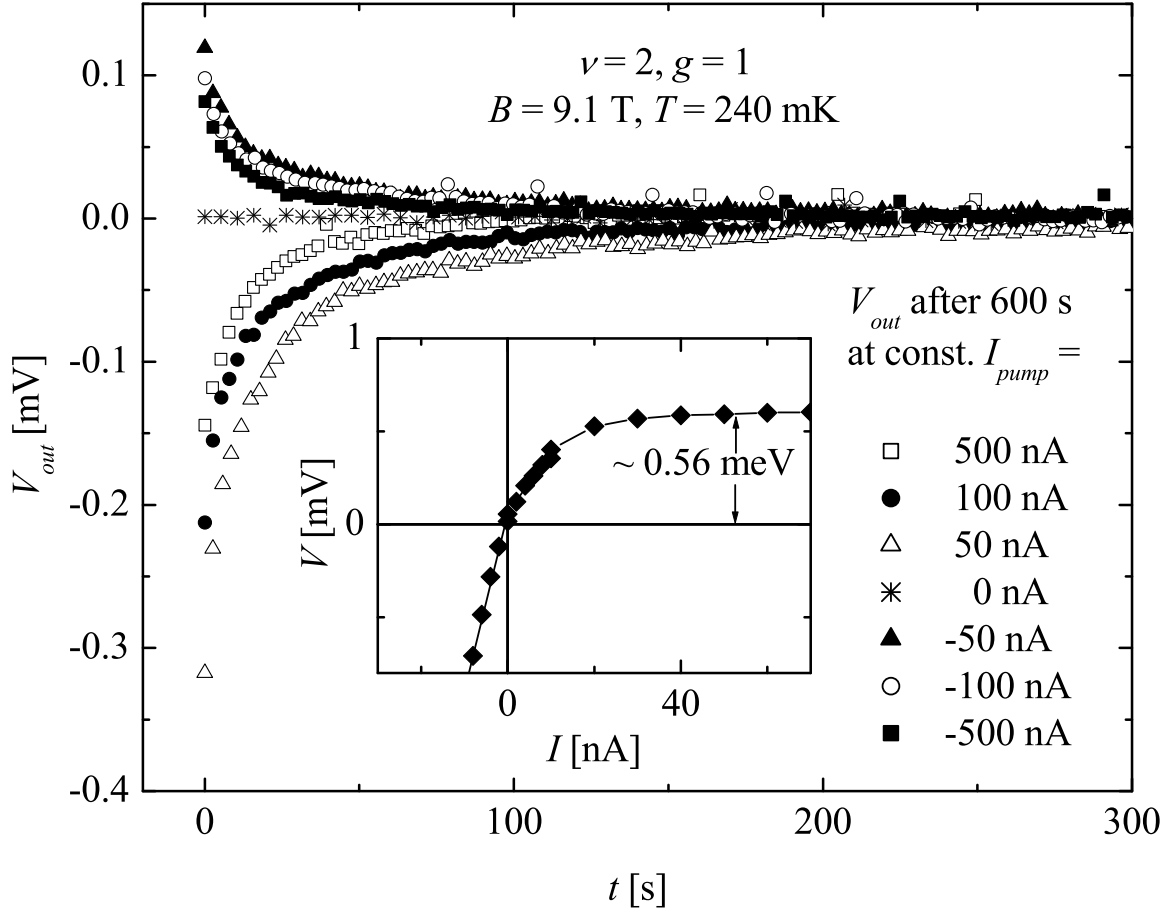}
\caption{Time-dependent output voltage between inner and outer
contacts after establishing a nuclear polarization for ten minutes
at different currents $I_{pump}$. The filling factors are $\nu=2$ 
and $g=1$ in the ungated and gated regions, respectively. The inset 
shows the non-linear $I-V$-characteristic used to determine $\Delta E_{z}$ (see text).  \label{fig2}}
\end{figure}

When the DNP has been established and
$I_{pump}$ is switched off, the voltage between the inner and
outer contacts does not vanish. This is somewhat surprizing,
since it indicates that there is
a source of energy present in the sample. Figure
\ref{fig2} shows that the output voltage $V_{out}$ is dependent
on the magnitude of
$I_{pump}$ in a non-monotonic fashion and that it exhibits a slow
exponential decay as a function of time. From the
dependence of the output voltage on $I_{pump}$ (both sign and magnitude),
a number of spurious effects can be excluded as explanations for this
striking observation: Thermal effects, e.g., caused by local heating of
the sample should be independent of the current
direction. A
transient charging of the highly resistive bulk region of the 2DEG
should exhibit a positive $V_{out}$ for a
positive $I_{pump}$.

\begin{figure}
\includegraphics[width=\columnwidth-2cm]{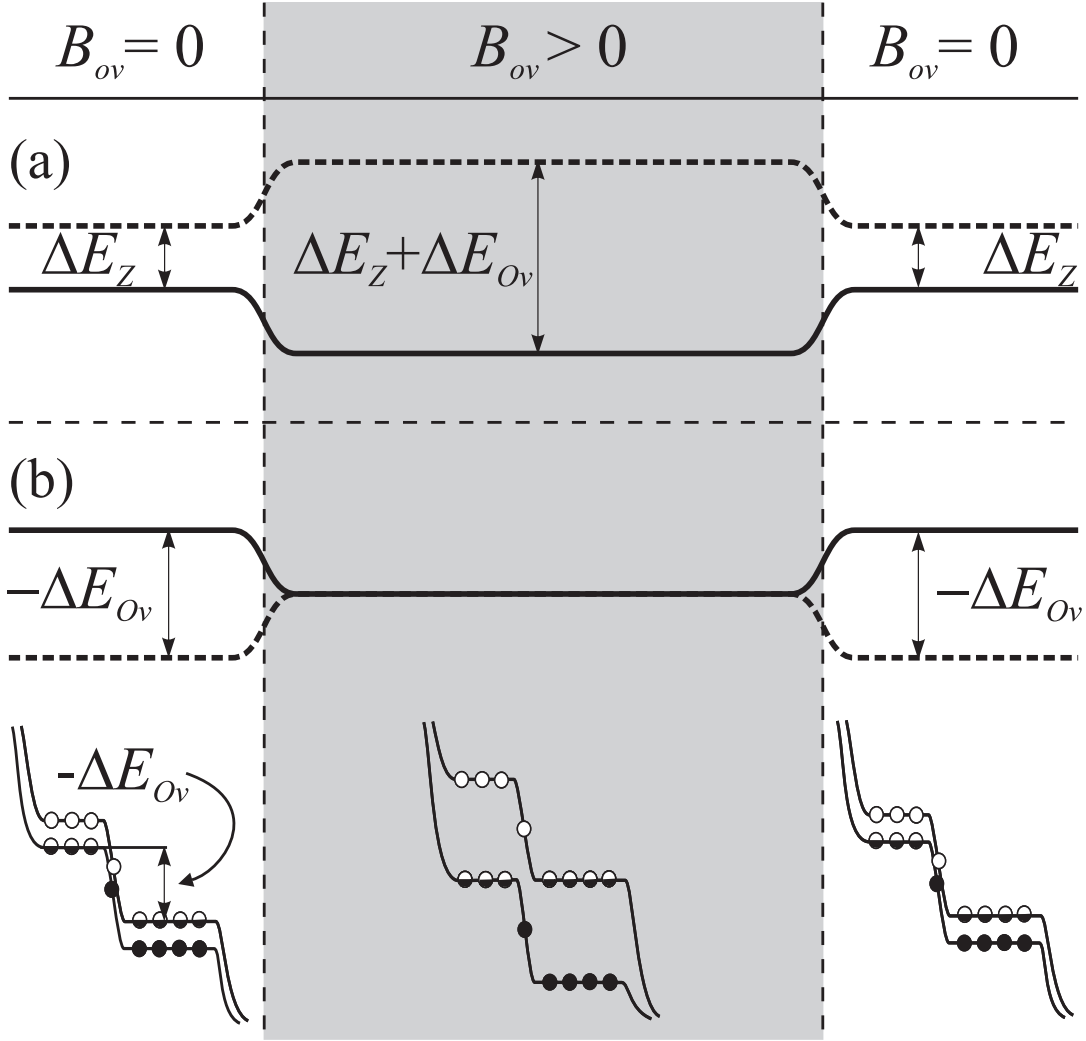}
\caption{Schematic representation of the local energy structure of
(a) non-interacting electron states and (b) edge states in local equilibrium.
The grey area
indicates the gate gap region with polarized nuclear spins, where an
Overhauser field $B_{ov}>0$ is present. The bottom inset shows the
edge reconstruction inside (center) and outside (left and right) of
the gate gap region. Half-filled circles represent electronic
states at the Fermi energy.\label{fig3}}
\end{figure}

On the other hand there is strong evidence that the internal source of
energy which gives rise to the output voltage is related to the nuclear
spin system. The output voltage exhibits a time dependence that
is similar to the one observed when the nuclear polarization is
established \cite{edik}, with a decay time that is typical for nuclear
depolarization \cite{dixon,wald}. Relaxation of the nuclear
spin system is expected to be most efficient at the sample edge.
The reconstruction of the edge potential \cite{shklovski} results in a
high density of electron states at the Fermi energy and provides for
a continuous energy dispersion, so that energy conservation is easily fulfilled
during flip-flop scattering (cf. also the
discussion of Fig. \ref{fig3}, below). It is therefore expected
that the decay of the nuclear polarization will take place by an
inverse flip-flop process and that this process will lead to a population
imbalance between the
spin-polarized edge states. At first glance, however, it seems impossible
that this  process can generate a voltage in the millivolt range, since the
energy gain from a nuclear spin-flip is approximately three orders of
magnitude smaller.

To account for our experimental observation, we will
introduce in the
following a simple model, based on the local energy shift of the
spin-resolved edge states. The required energy of a few $100\:\mu$eV is
supplied by the interaction energy between electrons and
nuclei rather
than the nuclear Zeeman energy alone. Loosely speaking, the energy is taken
out of the Overhauser field.

Figure \ref{fig3}(a) shows a sketch of the local energy for the spin-down
(dashed line) and the spin-up electron system (solid line) as it would develop 
if the local spin-polarization was present in the bulk of the sample:
In the unpolarized regions (left and right side of the figure), the energy
gap between the different spin systems corresponds to the bare Zeeman
splitting $\Delta E_{z}=|g^{\ast}|\mu_{B}B$, where $g^{\ast}$ is
the effective electronic Land\'{e}-factor (including exchange effects).
In the region where nuclear spin polarization
is present (middle, shaded
area), the hyperfine interaction will cause an
additional Overhauser shift $\Delta E_{Ov}$.

In the presence of the edge potential, however, current-carrying edge channels 
will develop, which --in local equilibrium-- share a common
chemical potential (see bottom of Fig. \ref{fig3} for sketches of the
edge reconstruction in the representation given by Chklovskii \textit{et al.}
\cite{shklovski}). In the present sample structure, the edge
channels can only interact in the very same region where the nuclear
polarization is present (hatched area in Fig. \ref{fig1}). Therefore, 
the energies of the two spin systems will be aligned, but only inside the gap 
regions \cite{minute}. Everywhere else,
particularly at the location of the voltage probes, the shift by $-(\Delta E_{z}+\Delta
E_{Ov})$, induced by the equilibration in the gate gap region, will
result in an energy difference of $-\Delta E_{Ov}$ (see Fig. \ref{fig3}(b)).

To understand the polarity of the output voltage with respect to
the sign of the pumping current, we analyze how $I_{pump}$ affects
the Overhauser field $B_{ov}$. Assuming the inner current contact to be
grounded, a positive $I_{pump}$ corresponds to electrons flowing
from the inner to the outer edge channel. This requires an electron
spin-flip from down to up. The corresponding nuclear spin-flop
from up to down induces a net nuclear `down' polarization
($\langle I_{z}\rangle < 0$)
which in turn results in a positive $B_{ov}$ (due
to the negative sign of $g^{\ast}$)
\cite{dixon,edik}. Accordingly, a positive $I_{pump}$ will lead to an
increased energy splitting $\Delta E_{z}+\Delta E_{Ov}$ between
the spin-polarized edge states, as illustrated in Fig. \ref{fig3}. Thus,
the output voltage, measured between the two edge channels will be
\textit{negative}, $V_{out}=-\Delta E_{Ov}/|e|$.

\begin{figure}
\includegraphics[width=\columnwidth]{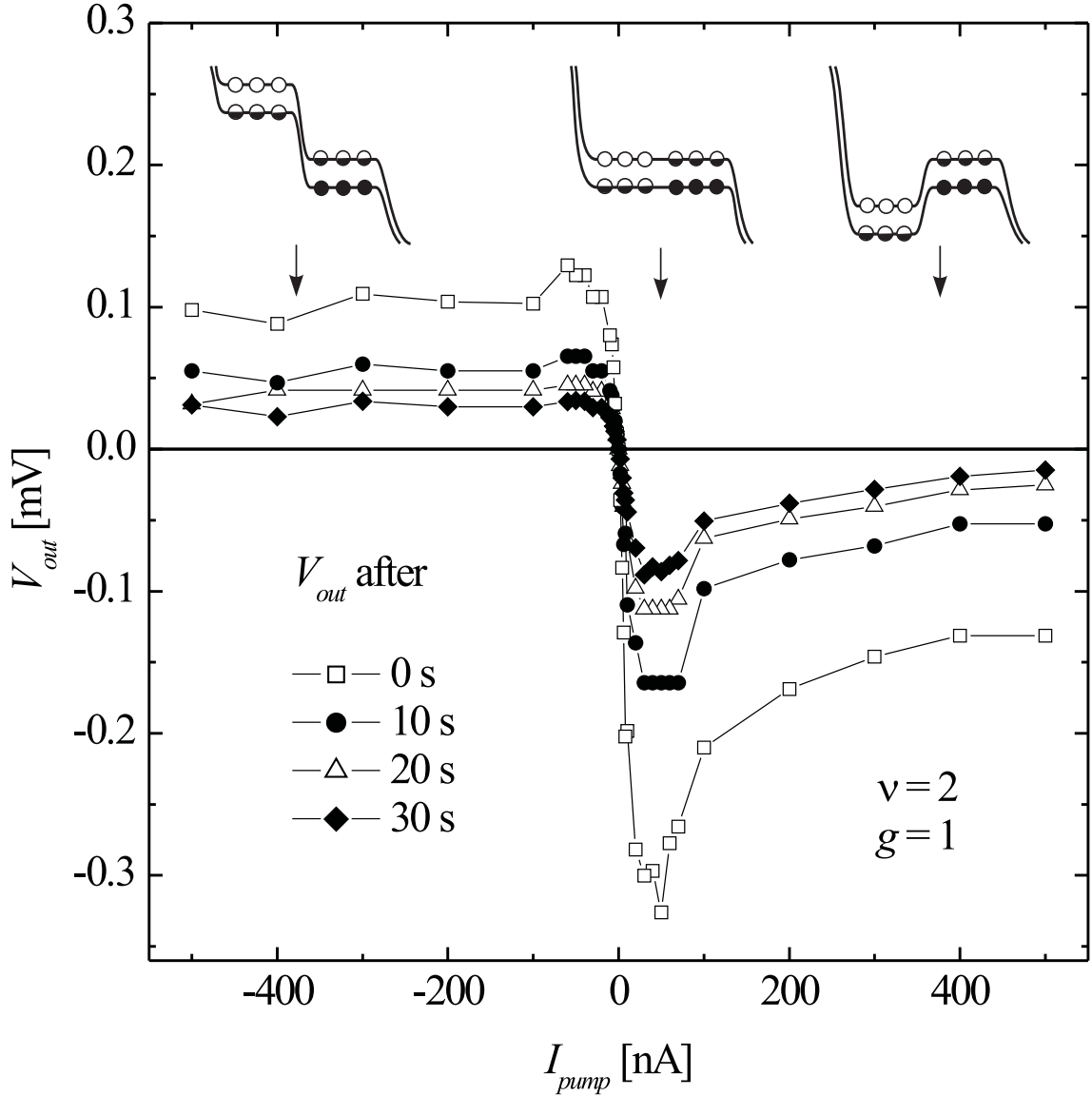}
\caption{Output voltage $V_{out}$ as a function of the pumping
current, for different times after switching off $I_{pump}$.
Pumping is most effective around $I_{pump}=+50$ nA. At this
current the resulting bias allows for resonant electron transfer
between the different spin systems, as schematically shown in the
middle inset. \label{fig4}}
\end{figure}

Our experiment therefore gives us direct access to the Overhauser shift in the
region of local nuclear spin polarization. Figure \ref{fig4} shows in
more detail the measured output voltage as a function of $I_{pump}$.
It can be seen that maximum $|V_{out}|$ is observed around $I_{pump}=+50$
nA and that the Overhauser field does not saturate but rather
\textit{decreases} for higher pumping
currents. Already for 300 nA, the
polarization mechanism is only half as effective  as for
$+50\:\mathrm{nA}$. This decrease in the nuclear
polarization with increasing positive bias ($I_{pump} > 100$ nA)
is in good agreement with previous results on a similar sample,
obtained using a different experimental
approach \cite{edik}.

A possible explanation for the sharp maximum
in $|V_{out}|$ is sketched in the insets to Fig. \ref{fig4}. Here,
the energetic structure of the edge states \cite{shklovski} is shown for a
negative (left), small positive (middle) and large positive
(right) bias voltage  which corresponds to the current $I_{pump}$
between the edge states \cite{alida}. For both negative and large positive voltage, the
electrons have to undergo tunneling processes and/or inelastic
scattering to transfer from one channel to the other. When the applied voltage is equal to the energy gap between the spin
resolved states, electrons can resonantly transfer between the inner and
outer edge channels, which is accompanied by a dramatic decrease in the resistance \cite{alida} (see inset in Fig. \ref{fig2}). This greatly 
enhances the effectiveness
of the nuclear spin polarization in the gate gap and leads to the
observed maximum in $|V_{out}(I_{pump})|$. A possible explanation for
this behavior is that the inelastic processes necessary for
equilibration at negative and high positive are so slow that the
relaxation (and thus the flip-flop process) takes place outside of
the gap region, where it does not contribute to the nuclear
polarization in the gate gap.
Further experiments are necessary, however, to
clarify this point.

The results shown in Fig. \ref{fig4} can be used to estimate the
maximum value of the Overhauser-field in the gate gap:
$B_{ov}=B|eV_{out}|/\Delta E_{z}$ \cite{Bov}, where $B=9.1$ T is the
external field and $\Delta E_{z}$ is the exchange-enhanced
electronic spin splitting. The latter can be determined to be 
$\Delta E_{z}\approx 0.56$ meV by
evaluating the non-linear $I-V$-curve \cite{alida}, shown as an
inset to Fig. \ref{fig2}. To ensure that $\Delta E_{z}$ is free of contributions 
from the hyperfine field, the nuclear spins are allowed to relax for $600$ s at 
$I_{pump}=0$ before each measurement $V(I)$ is taken. 
With $|eV_{out,max}|\approx 0.32$ meV and 
$\Delta E_{z}\approx 0.56$ meV we 
find $B_{ov,max}= 5.2$ T. This
is  among the highest values determined experimentally so
far and close to the maximum Overhauser field of 5.3 T, predicted for
100\% nuclear polarization in GaAs \cite{paget}.
We attribute this high degree of nuclear polarization to the
optimum choice of the pumping current which corresponds to a
resonant electron transfer between the spin split states (see Fig.
\ref{fig4}). 

Finally, we would like to mention that the
polarized nuclear
spin system in the present device not only serves --via the hyperfine interaction--
as an ordinary, battery-type
voltage source. Because of the 100\%
spin-polarization of the edge channels, a \emph{spin current} will be concomitant with
any charge current drawn from our device. In this
respect, the present experimental realization resembles the various schemes
for 'spin batteries' (with and without electric
potentials) that have been envisioned  recently \cite{brataas}.

\begin{acknowledgments}
The present work is based on an ongoing and fruitful collaboration
with Valeri T. Dolgopolov and Eduard V. Deviatov. We would like
to thank Alexander O. Govorov and Wiebren S. Veeman for valuable
discussions. Financial support by the Deutsche
Forschungsgemeinschaft, SPP "Quantum Hall Systems", under grant LO
705/1-3 is gratefully acknowledged.
\end{acknowledgments}


\begin{thebibliography}{99}
\bibitem{minute} Note that during the decay of the nuclear
polarization, a minute difference in chemical potential may be
present, since the inverse flip-flop process will cause an
imbalance between the electron spin systems.
\end{thebibliography}
\end{document}